\documentclass[twocolumn]{aa}
\usepackage[dvips]{graphicx}
\usepackage{parskip}
\usepackage{txfonts}
 
\begin{document}

\title{SIMBA's view of the $\epsilon$~Eri disk\thanks
      {Based on observations collected at the European Southern
       Obser\-vatory, La~Silla, Chile (71.C-0001)}}

\author{O.~Sch\"utz
        \inst{1}
        \and
        M.~Nielbock
        \inst{2}
        \and
        S.~Wolf
        \inst{3}
        \and
        Th.~Henning
        \inst{1}
        \and
        S.~Els
        \inst{4}
        }

\offprints{schuetz@mpia.de} 

\institute{Max-Planck-Institut f\"ur Astronomie, K\"onigstuhl 17,
           D-69117 Heidelberg, Germany
           \and
	   Ruhr-Universit\"at Bochum, Astronomisches Institut, 
	   Universit\"atsstr. 150, D-44780 Bochum, Germany 
           \and
	   California Institute of Technology, 1201 E California Blvd,  
	    Mail code 105-24, Pasadena, CA 91125, USA
	   \and
           Isaac-Newton-Group of Telescopes, Apartado de Correos 321,
           38700 Santa Cruz de La~Palma, Spain
           } 

\date{Received 6 November 2003; accepted 5 December 2003} 

\abstract{

We present the first observational confirmation for an extended
circumstellar dust disk around \object{$\epsilon$~Eri}. The
observations were obtained with the bolometer array SIMBA at the 15\,m
radio telescope SEST in Chile and measure the dust continuum at
1.2~mm. The emission, with a total flux of 21.4~mJy and a rms of 
2.2~mJy/beam, is resolved to a deconvolved size of 27\farcs4 which 
corresponds to 88~AU. No clear indication for a ring-like disk
structure is seen, possibly also due to the telescope's large beam width of
24$''$. Models of the object's spectral energy distribution from IR to 
mm-wavelengths show that the emission can also be explained by 
a simple disk model. We further demonstrate the strong influence of noise 
and propose to be cautious with interpretations of the ring
substructure.

 
         }

\maketitle

\section{Introduction}

Circumstellar disks are an ubiquitous product of the formation of
stars and planetary systems. In these disks planets are believed to
form from planetesimals that in turn are created through coagulation
of dust and gas (Beckwith et al.\ \cite{Beckwith}). The final stages
of this process remain to be characterised but observations suggest
that previous protoplanetary accretion disks turn into debris disks
at the end of planet formation.

Photometric measurements with the IRAS satellite (Aumann \cite{Aumann})
provided first hints for dust around the $\sim$800~Myr old star
\object{$\epsilon$~Eri}, a K2V dwarf at 3.22~pc distance to the
Sun. From varying sub-mm flux densities in single-beam observations 
with different beam sizes (Chini et al.\ \cite{Chini90}, Chini et 
al.\ \cite{Chini91}, Zuckerman \& Becklin \cite{Zuckerman}), an extended 
-- possibly ring-like -- disk structure was assumed. In SCUBA
observations at 850~$\mu$m Greaves et al.\ (\cite{Greaves}) found the dust
emission peaking in a ring-like distribution between 35--75 AU
distance from the star. The central cleared region and an asymmetrical
substructure in the ring were explained by the influence of orbiting
planets. Now a planet with a semi-major half axis of 3.4~AU is known
(Hatzes et al.\ \cite{Hatzes}), which for more than a decade was
controversial because of the star's high chromospheric
activity. Quillen \& Thorndike (\cite{Quillen}) explained the ring
substructure with a further hypothetical planet at 40~AU distance. CO
line observations at the Swedish-ESO Submillimeter Telescope (SEST) 
concluded that the disk of 
\object{$\epsilon$~Eri} is likely devoid of any gas (Liseau
\cite{Liseau}). With ISO no features from dust were found in mid-IR
spectra between 6 and 12~$\mu$m, and in maps at 60 and 90~$\mu$m
no resolved emission was seen (Walker \& Heinrichsen \cite{Walker}). A 
partially resolved region in IRAS data, quoted by some authors, was
thus not confirmed. Several attempts to reveal the disk (in 
scattered light) with coronographic imaging at near-IR and optical 
wavelengths have been tried without success, including HST
observations. Recently, Li et al.\ (\cite{Li}) calculated a disk
model from the IR to sub-mm spectral energy distribution (SED) of 
\object{$\epsilon$~Eri}, but different disk and grain parameters were
used than in our simulations. Furthermore, our calculations now also 
include observations at 1.2~mm. An overview of the fluxes used in
this work is given in Table\,\ref{table:fluxes}. 

\begin{table}[b]
  \centering
  \caption{Fluxes of $\epsilon$~Eri used in this work. The mid-IR IRAS 
           fluxes contain also contributions from the stellar
           continuum. In the model SEDs in Fig.\,\ref{figure:sed} this
           is separated from the dust emission.}
  \begin{tabular}{|cccc|}
  \hline
  \hline
  Wavelength [$\mu$m] & Flux [mJy] & Error    &  Reference      \\
  \hline
  12                  & 9520       &   4\,\%  &  IRAS           \\
  25                  & 2650       &   6\,\%  &  IRAS           \\
  60                  & 1660       &   8\,\%  &  IRAS           \\
  100                 & 1890       &   9\,\%  &  IRAS           \\
  450                 &  185       & 103 mJy  &  Greaves (1998) \\
  850                 &   40       &   3 mJy  &  Greaves (1998) \\
  1200                & 21.4       & 5.1 mJy  &  this work      \\
  \hline
  \end{tabular}
  \label{table:fluxes}
\end{table}

We present the first observational confirmation of an extended
dust disk around $\epsilon$~Eri found with the Sest IMaging Bolometer 
Array (SIMBA). With results from numerical simulations for the target's 
SED we set constraints on dust properties and discuss whether a continuous 
disk with an inner radius given by the dust sublimation temperature could 
be considered alternatively.

\section{Observation}

The observations of $\epsilon$~Eri were obtained in July 2003 with the
37-channel bolometer array SIMBA at the SEST in La~Silla
(Chile) using the fast scanning mode at 250~GHz ($\lambda = 1.2$~mm). 
Skydips were taken every 2 to 3 hours to correct for the
atmospheric zenith opacity, which was between $\tau_0 \approx 0.23$
and $\tau_0 \approx 0.37$. We followed the SEST pointing model (rms $\sim$ 
2\farcs5) and checked the telescope pointing regularly. No significant 
deviation from the pointing model was seen. The flux calibration is based
on observations of Uranus. In total, 180 maps of $\epsilon$~Eri were 
integrated which correspond to a total on-source time of $\sim$15 hours.

\begin{figure*}[t]
 \centering
 \includegraphics[angle=270,width=\textwidth]{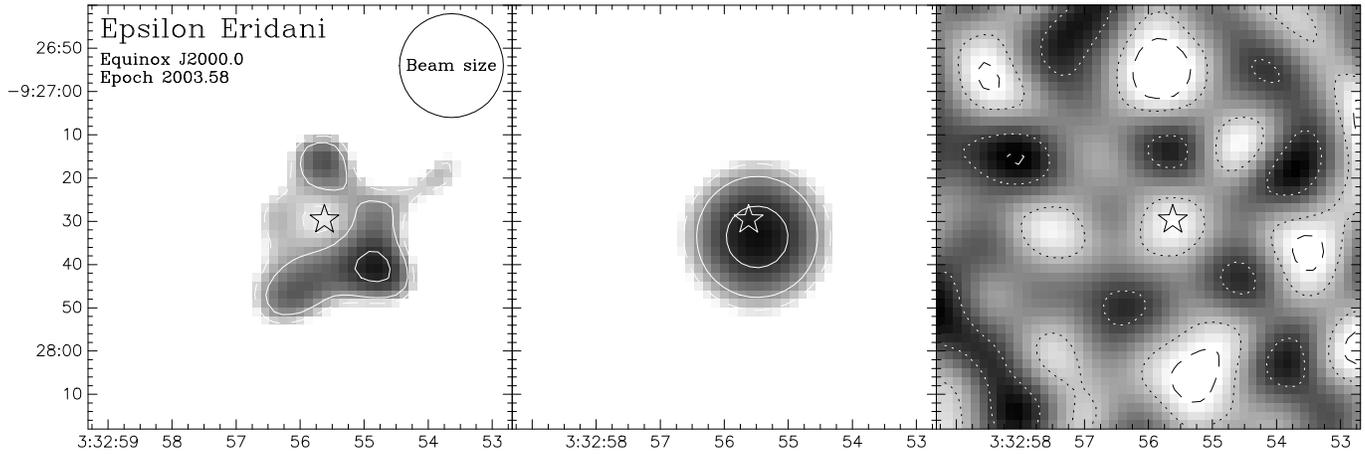}
 \caption{Similar to the SCUBA observation our image sampling is
          2$''$. We find a central cavity and the emission peaking in
          certain spots along a ring-like distribution (left panel). 
          However, subtracting a circular Gaussian intensity
          distribution with a FWHM of 36\farcs4 (central panel) only
          leaves residues of pure noise (right panel). Thus we conclude
          that the substructure in our map was likely caused by
          remnant noise. The star marks the location of $\epsilon$~Eri,
          corrected for its proper motion (-976.36 x 17.98 masec/year,
          cf. the SIMBAD database or Perryman et al.\ \cite{Perryman}). 
          In the left and central
          panel the intensity is scaled logarithmically from 5 to 
          10~mJy, solid contours represent regions with an accuracy of 3 
          resp. 4\,$\sigma$ (rms = 2.2~mJy/beam), while the dashed 
          contour (which nearly conincides with the outer margin of the 
          greyscaled region) corresponds to 2.5\,$\sigma$. The right panel 
          shows -2\,$\sigma$ resp. +2\,$\sigma$ as dashed contours and 
          $\pm$1.5\,$\sigma$ are symbolised by dotted lines.}
 \label{figure:obs}
\end{figure*}

Data reduction was performed with MOPSI\footnote{MOPSI has been
developed and is maintained by R.~Zylka, IRAM, Grenoble, France}
and included despiking, baseline fitting,
suppression of the correlated sky noise, opacity and gain-elevation
correction as well as co-adding the 180 single maps to a final one,
which is shown in Fig.\,\ref{figure:obs}. The residual root mean square
error (rms) amounts to 2.2~mJy$/$beam.

Observing at 1.2~mm with the 15~m dish of the SEST results in a nominal
beam size of 24$''$. This value can slightly vary with the elevation 
and other authors occasionally report 23$''$ or 25$''$. In principle, 
the beam size can be derived from a point source like, e.g., the planet 
Uranus. Averaging over various telescope positions we obtained a beam 
width of 24\farcs2 $\pm$ 0\farcs2. The resolved emission of 
\object{$\epsilon$ Eri} can be fitted by an extended, circular Gaussian 
intensity distribution with a full width half maximum (FWHM) of 36\farcs4. 
This fit and its centering is achieved with a routine in MOPSI. Relative
to the beam width we resolve the dust emission around $\epsilon$~Eri
consistent with a deconvolved total extent of 27\farcs4 corresponding 
to 88~AU. An elliptical shape of the intensity distribution -- indicating
an inclination of the disk -- cannot be excluded, but is not apparent 
from our measurements. Greaves et al.\ (\cite{Greaves}) reported 
$i$~=~25\degr. At 1.2~mm, after subtracting a photosphere contribution 
of 0.9~mJy, the total disk flux amounts to 21.4~mJy with a peak flux of 
9.7~mJy/beam and 2.2~mJy/beam noise. As can be seen in 
Fig.\,\ref{figure:obs} (central panel), the peak of the Gaussian lies 
7$''$ southwest of the star. This is not necessarily a physical effect,
since the accuracy of the centering routine is within 5$''$ (corresponding 
to 2.5 pixel in this figure).

\section{Numerical simulations}

The disk model is derived from the target's SED with the fluxes shown in 
Table\,\ref{table:fluxes}. Since especially the mid-IR fluxes contain
a contribution from the stellar continuum, this is treated separately
in the simulated SEDs.

We apply a model of an optically thin disk with the simple radial density 
profile $n(r) \sim r^{-1}$ (for the calculation of the dust temperature 
distribution see Wolf \& Hillenbrand \cite{Wolf}, Eq.~5). The following 
stellar parameters are used for $\epsilon$~Eri: $T_\star = 4700$~K, 
$R_\star = 0.8~R_\odot$ and $L_\star = 0.33~L_\odot$ (e.g. Soderblom \& 
D\"appen \cite{Soderblom}). We select a circular disk with an extension 
from $R_{\mathrm{in}} \approx 0.025$~AU (corresponding to the radius given 
by the dust sublimation temperature) up to the observed 
$R_{\mathrm{out}} = 88$~AU. Other choices for the inner disk radius and 
their consequences are discussed in the next chapter. 

\begin{figure}[t]
  \centering
  \includegraphics[scale=0.49]{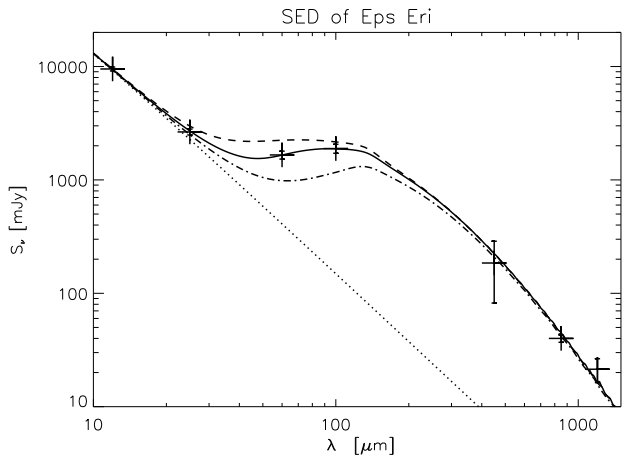}
  \caption{Modelled SED of $\epsilon$~Eri. The simulations were
           calculated with a grain size distribution and varying
           minimum grain size: $a_{\mathrm{min}} = 4 \,\mu$m (dashed 
           line), $a_{\mathrm{min}} = 16 \,\mu$m (dash-dotted), 
           $a_{\mathrm{min}} = 8 \,\mu$m (solid curve, best fit). The 
           stellar photosphere is shown as a dotted line.}
  \label{figure:sed}
\end{figure}

Spherical homogeneous dust grains were used with radii $a$ according to a
size distribution $n(a) \sim a^{-3.5}$. This describes the equilibrium 
size distribution resulting from a collisional cascade of dust (Tanaka 
et al.\ \cite{Tanaka}) and has often been used for the interstellar medium 
(ISM) and modelling of circumstellar disks. The best approximation to the 
observed SED is obtained for a minimum grain size 
$a_{\mathrm{min}} = 8 \,\mu$m which is displayed in Fig.\,\ref{figure:sed}. 
SEDs for $a_{\mathrm{min}} = 4 \,\mu$m and $a_{\mathrm{min}} = 16 \,\mu$m 
are shown for 
comparison. We use an upper grain size of 1~mm. The disk mass, given in 
the next section, refers to this upper size limit, since we have no direct 
observational hints for cm-sized grains. Simulations are performed with a 
dust mixture deduced from the ISM: 62.5\% silicate grains (according to 
Draine \& Lee \cite{Draine} and Weingartner \& Draine \cite{Weingartner}) 
mixed with 37.5\% graphite. The dust grain density 
amounts to $\rho$(silicate) = 2.7~g~cm$^{-3}$ and 
$\rho$(graphite) = 2.24~g~cm$^{-3}$. 

\section{Discussion}

{\bf Disk temperature and dust mass:}

Submillimeter observations probe the cold dust towards the outer part of the
disk. From a fit to the observed fluxes between far-IR and mm-wavelenghts
we determine for the cold dust component a temperature of $50$~K using 
\,$\beta$ = 0 and $\kappa$(1.2~mm) = 1.7~cm$^2$g$^{-1}$ (a value often used 
for Vega-type disks, e.g.\ Sylvester et al.\ \cite{Sylvester}). With the
numerical simulations described above we obtain a dust mass of $5.2 \times 
10^{-9} M_\odot$ = $1.7 \times 10^{-3} M_\oplus$, which lies about a factor 
of three under the lower mass limit given in Greaves et al.\ (\cite{Greaves}).

{\bf Dust ring or a continuous disk:}

We are able to explain the observed SED with a simple disk model, an 
inner disk radius given by the dust sublimation temperature and a 
minimum of free parameters, i.e. without further assumptions. However, 
we do not exclude other disk models. With the limited resolution of
the SEST beam only a very sharp ring would have been detectable. Even
for SCUBA the reported central cavity is close to the resolution
limit. To explain however the mid-IR IRAS fluxes, some amount of hot
circumstellar dust must be present at regions relatively close to the 
star. From modelling the observed SED no statement can be made to which 
extent the inner disk region may be depopulated of dust. Due to the
large parameter space in any dust disk model (disk density profile, 
dust grain radii, grain size distribution, dust mineralogy, grain shapes, 
etc.) several solutions may exist to interpret the observed SED. Li et 
al.\ (\cite{Li}) find their model in good agreement with Greaves et 
al.\ (\cite{Greaves}) for disk radii $\geq$~28~AU, while an additional 
``zodiacal'' dust component would be required to explain the possible 
emission from the inner parts of the disk. This inner region was not 
clearly resolved by previous mid- and far-IR satellite cameras, but
can be better studied with upcoming SIRTF observations (see Li et 
al.\ \cite{Li} for a discussion on this). An alternative approach would 
be interferometric observations with the IRAM Plateau de Bure 
Interferometer operating at 1.3~mm wavelength with a resolution up to 
$1''$.

{\bf Influence of noise on the observed disk substructure:}

Fig.\,\ref{figure:obs} (left panel) shows our final co-added map of
\object{$\epsilon$~Eri}. The pixel sampling with 2$''$ is justified
because of SIMBA's image sampling of 15\,ms in azimuthal scan
direction. With 80$''$/s scanning speed this results in a spatial 
sampling of 1\farcs2. In right ascension and declination we obtain a 
similar value due to
the large coverage of maps taken at different phase angles. Our final 
image resembles a disk with a cleared central region and substructures in 
the ring. However, we can show that these details are most likely
caused by remnant noise. The resolved emission of \object{$\epsilon$
Eri} can be fitted by an extended Gaussian intensity distribution with
a FWHM of 36\farcs4 (central panel). A subtraction of this fit from
the final map only results in pure residual noise, while a real
central cavity should cause a clear negative peak after subtraction. 
The occurence of a minimum exactly at the position of the star leaves 
speculations whether this is a contribution of a real gap. However, 
this minimum is even less prominent than the noise minima above and below 
the star's position. 
Similarly, the knots in the outer disk part can be explained by 
noise maxima. To verify this, we simulated a ring-like disk with 
artificial noise but without intrinsic ring substructures, and obtained a
comparable result. 

The knots in the SCUBA ring were attributed by some 
authors to be traps of dust due to orbital resonances of Jovian planets. 
However, most of these are smaller than the resolution of the SCUBA
beam. We speculate that these structures may be explained similar to
our observations.

\section{Conclusions}

Since several publications calculate planetary parameters from the
knots in the \object{$\epsilon$~Eri} ring structure, we intended to
perform the first confirmation of this disk shape. Our observation with
the bolometer array SIMBA operating at 1.2~mm is the first verification
of an extended disk around this star. However, these data do not provide 
clear indications for the previously reported ring substructure. In our
case the knots can be explained with remnant noise effects. We are able 
to model the observed SED with a simple, gap-free disk, but do not rule 
out the existence of a possible central cavity.

\begin{acknowledgement}

We thank the SEST staff for their helpful support. SE is
supported under Marie-Curie Fellowship contract HDPMD-CT-2000-5.

\end{acknowledgement}

\end{document}